# **Analysis on the relations between piano touch and tone**

A thesis presented by

Lai-Mei Nie

to

Department of Physics

in partial fulfillment of the requirements of the course

*General Physics*

Tsinghua University

# Analysis on the relations between piano touch and tone

Lai-Mei Nie

Department of Physics

Tsinghua University

Beijing 100084, China

May, 2008

Abstract: In piano playing, different ways of touch lead to different tones. The effects of changing forces on keys are presented by changing interaction between hammer and string. The thesis focuses on several important variables in hammer-string system and draws conclusions to the question about piano touch and tone.

Piano has existed for more than 600 years, from the harpsichord to the modern prototype built by the Italian Cristofori in the year of 1709, the instrument has been improved a lot in both mechanical structure and acoustical properties, with the efforts of manufacturers, pianists and piano technicians. Their great techniques, creation and abundant experience lead the art of piano to present prosperity.

The reason why piano attracts so many people is mostly because of its unique tones. As an amateur piano player, I find that a piano, can own such a variety of voices, from the soft and sweet ones in Mozart's sonatas, to the strongest and brightest in Chopin's Polonaises or Liszt's Rhapsodies. This spectacular characteristic of piano relies heavily on the experience of manufacturers and piano technicians, even though many parameters of piano have been measured to help improve its quality. There are indeed some experimental measurements to explain why a piano can sing so marvelously, and some theoretical analyses that refer to nonlinearity or more complex aspects. Here we try to give a possible and quantitative solution to the question without using too much complicated mathematical methods. More concretely, we'll deduce how the vibrations on a piano string vary with the intensity of force exerted on the key.

While the tone we hear partly depends on our subjective feelings, it is still imbued with a lot of physical phenomena and principles. Theories of acoustics and mechanics are used to quantitatively analyze the motion of piano string and the interaction between hammer and string (an illustration of piano structure: Page 2, Fig1.1).

We hope this research can offer help to manufacturers and piano technicians when they work. At the same time, there are still several open questions, but this encourages us to do more in the physical research about piano making and playing.

The thesis consists of the following parts:

1. Some concepts, prerequisites and approximations.
2. Experiments to confirm that a piano key can generate different tones.
3. Physical analysis, initial conclusion and discussion.
4. Corrections of the analysis, discussion and final conclusion.
5. Discussions of unsolved problems.
6. Acknowledgements.
7. Appendix: some mathematical details of Part 4.
8. References.

## 1. Some concepts, prerequisites and approximations

(I) We discuss the tone of prompt sound here. It starts almost as soon as the key is pressed, in distinction with the aftersound, which can last for a much longer time.[1]

(II) We discuss upright piano here. Upright pianos are widely used at home and in schools, while the grand pianos are used for concerts.

Here is an illustration of the keyboard mechanism in an upright piano.

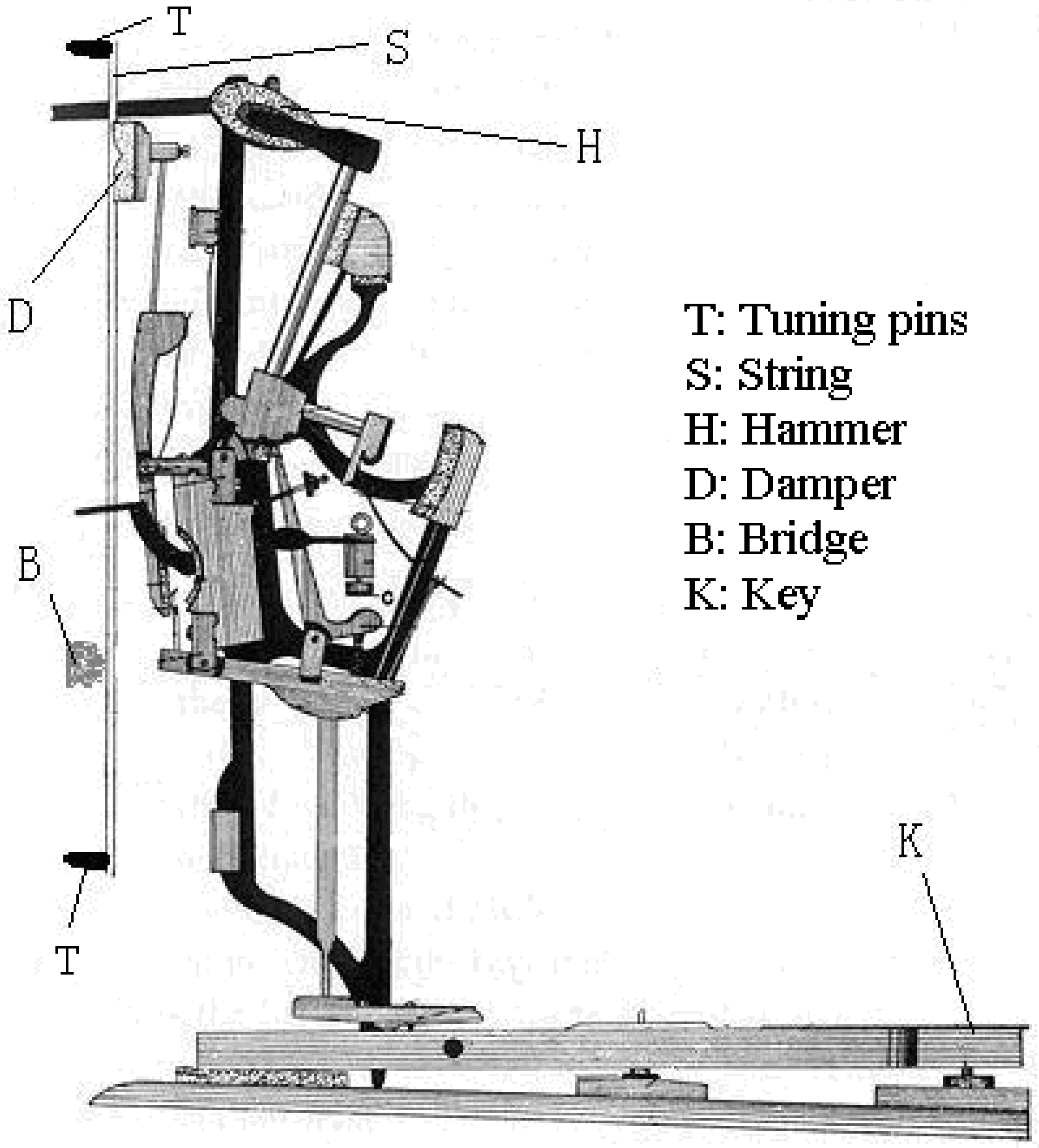

Fig 1.1. Keyboard mechanism of upright piano
(Amended from en. wikipedia. org/ wiki /Action_(piano))

When the key is pressed down, energy is transported through three levers and finally becomes the kinetic energy of the hammer. The hammer strikes the string to make sound. The damper lifts so that the string can vibrate freely.

The string is fixed at three points: two tuning pins and a bridge. The bridge is used to transmit string’s energy to the soundboard, which is able to vibrate with the air around it to make bigger sound. There are equipments to reduce the vibrations on the ineffective parts (from the lower “T” to “B” in Fig 1.1). Only the effective part (from the upper “T” to “B” in Fig 1.1) contributes to the tone.

The tuning pins are rigid supports while the bridge is not a complete rigid one,

otherwise it cannot transport the vibrations to the soundboard. However, since we study the prompt sound during which the string's energy loss at the bridge is of small amount and there are nails on the bridge to fix the string tightly, we can treat the bridge as a rigid support like the tuning pins. (More discussions in Part 5)

(III) Most of the keys correspond to two or three parallel and unison strings[1], and we study one of them. Since the hammer strikes the strings not exactly vertically, there are some twisted vibrations on the strings. But compared with the transverse vibrations, its effect can be neglected in our research.

(IV) Piano's tones depend on the ratios of fundamental and harmonic frequencies of the standing waves on the string. Bright sound contains more high frequencies while soft and dim sound contains more fundamental frequency. The intensities of the very high frequencies are too weak to affect the tone[2], so our research focuses in the range from the first to the sixth frequencies.

(V) Practically, when our fingers knock the keys more strongly, both the magnitude of the press force and the speed at which the key sinks are increased. In piano playing, it will be very difficult and of little sense to increase only the magnitude of the force without increasing the finger's pressing speed. So when we mention a bigger press force, we indicate that both the magnitude and the speed are increased.

## 2. Experiments to confirm that a piano key makes different tones under different forces

"The more strongly hit, the brighter sound a piano key will make." This is agreed by many people who have played or listened to piano. However, when the press force becomes bigger, the loudness increases, too. Therefore, there is possibility that the ratios of fundamental and harmonic frequencies on the string remain unchanged but due to the increased loudness (the increased SPL), people still think the tones are different.

To exclude the effect of loudness change, I put a voice recorder in appropriate distances with a piano to keep the SPL the same and recorded its sound when a key in the middle range was pressed with different magnitudes of forces. Then the recorded sounds were sent into an oscilloscope and a computer respectively to be changed into visual waves. Results are listed in Table 2.1.

| Forces (vertical) | Slight touch | Mighty touch |
|---|---|---|
| Tone | Dim and soft | Bright and sharp |
| Average SPL of prompt sound (software: Samsung Voice Manager) | -25dB | -25dB |
| Oscilloscope | Wave 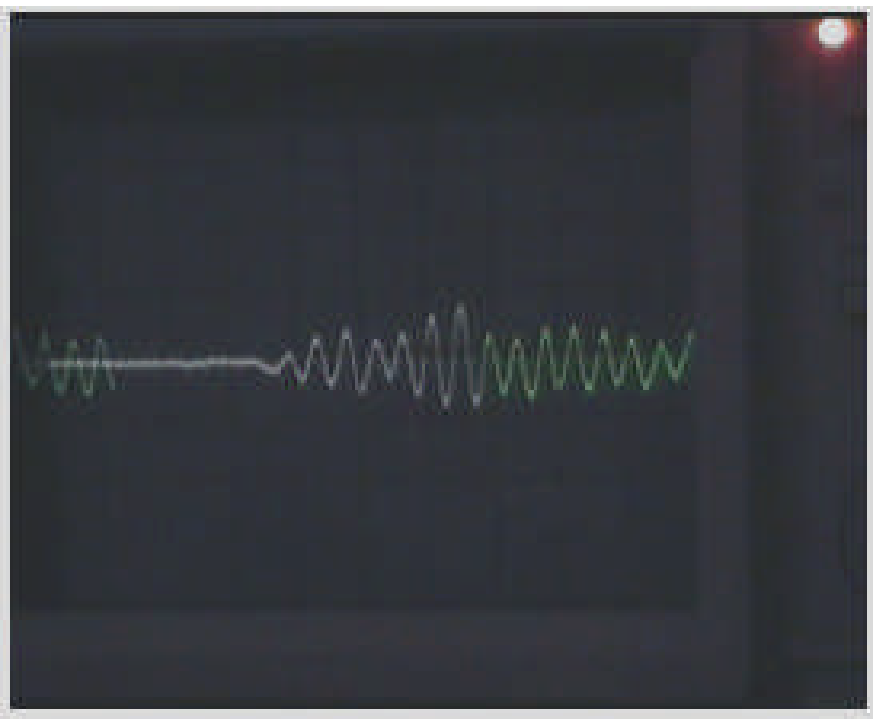 | Wave 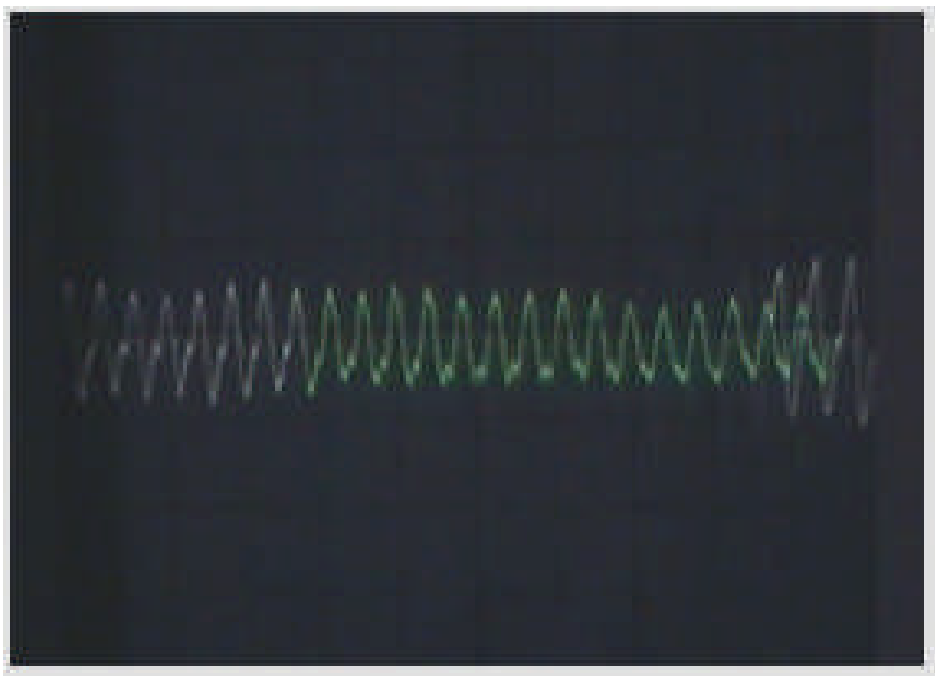 |
| Computer (software: Windows Media Player) | Wave 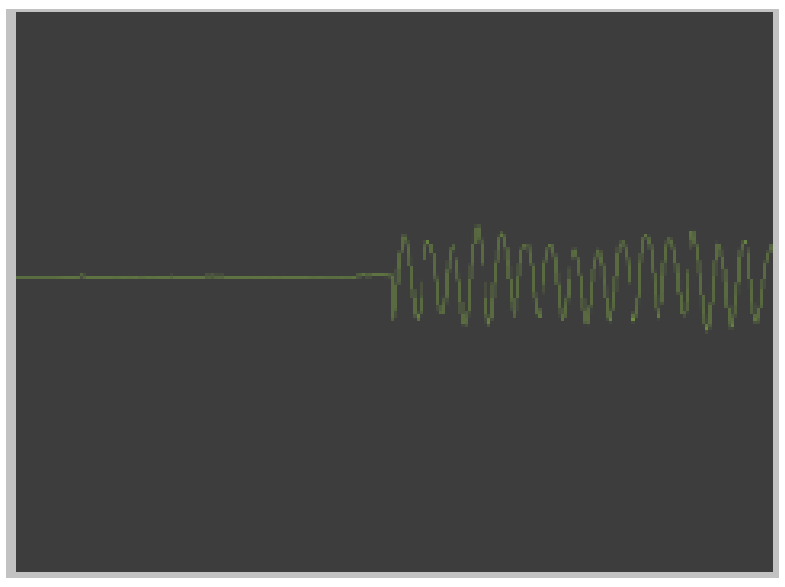 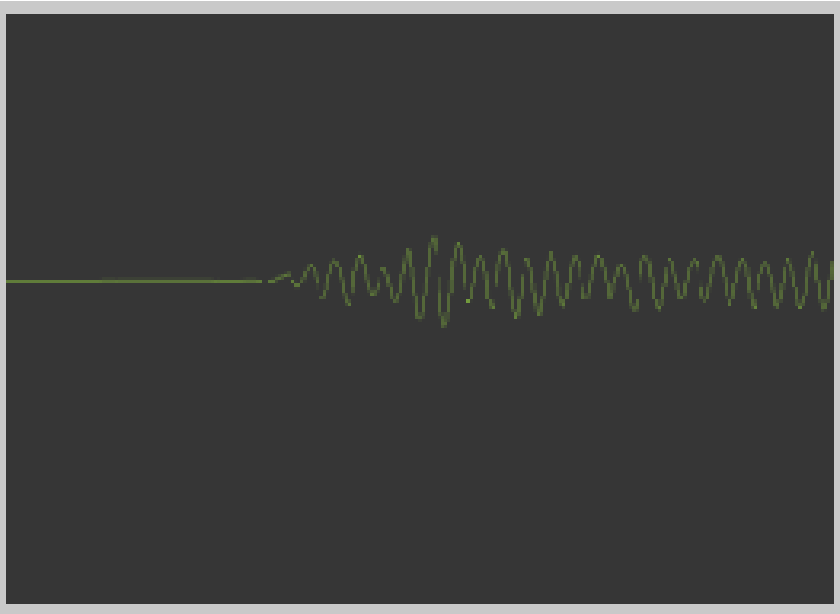 | Wave 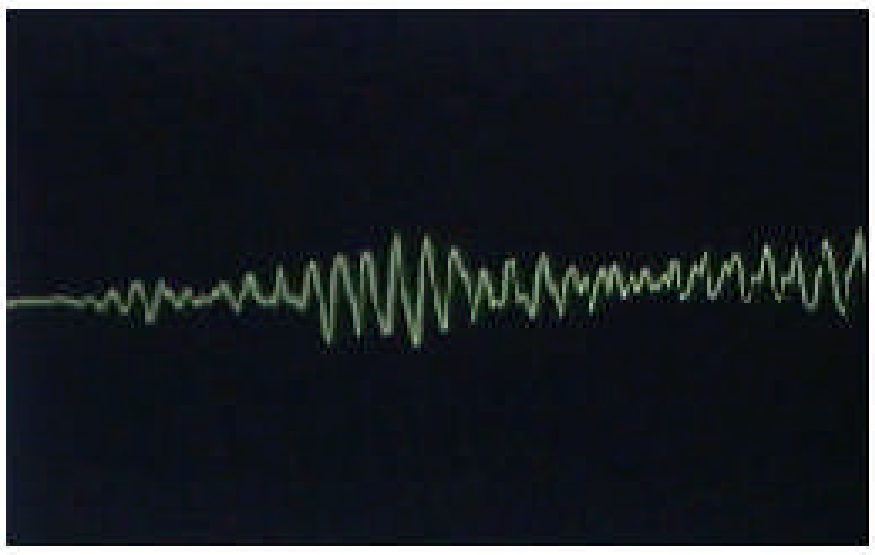 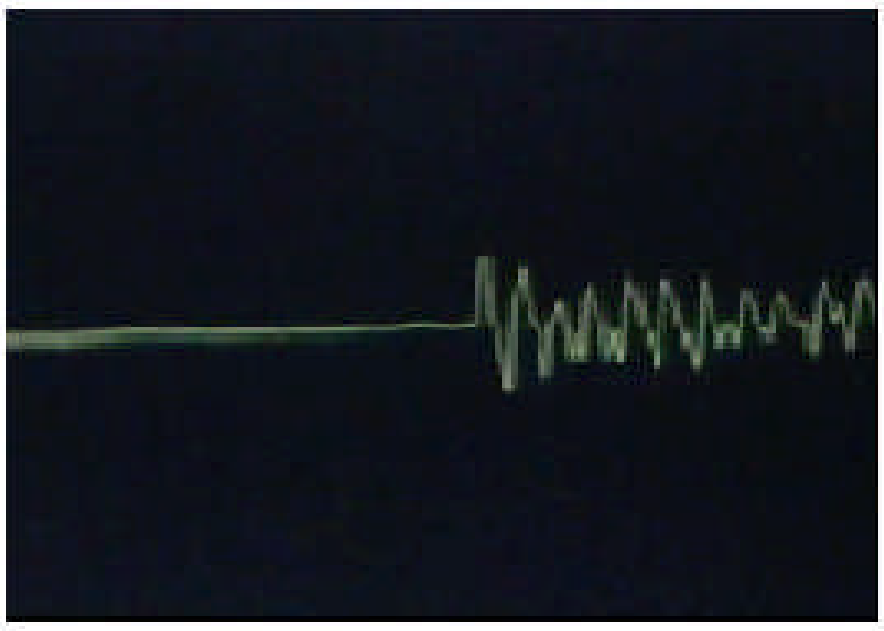 |
| Analysis | It's almost a sine curve. The fundamental frequency $f_1$ dominates. | It's the superposition of several frequencies, mainly $f_1$, $f_2$, $f_3$. |

Table 2.1 Experimental verification of different tones resulted from different touches. Because of the computer's manipulation, the SPL here is not the real SPL of sounds, but can be used for comparison.

With the effect of loudness excluded, it's confirmed that when a key is under different pressures, the ratios of the intensities of frequencies on the string are indeed different.

## 3. Physical analysis, initial conclusion and discussion

We derive wave equation on a vibrating string under the condition of small amplitudes. Set $u(x,t)$ to be the displacement of point $x$ on the string at time $t$.

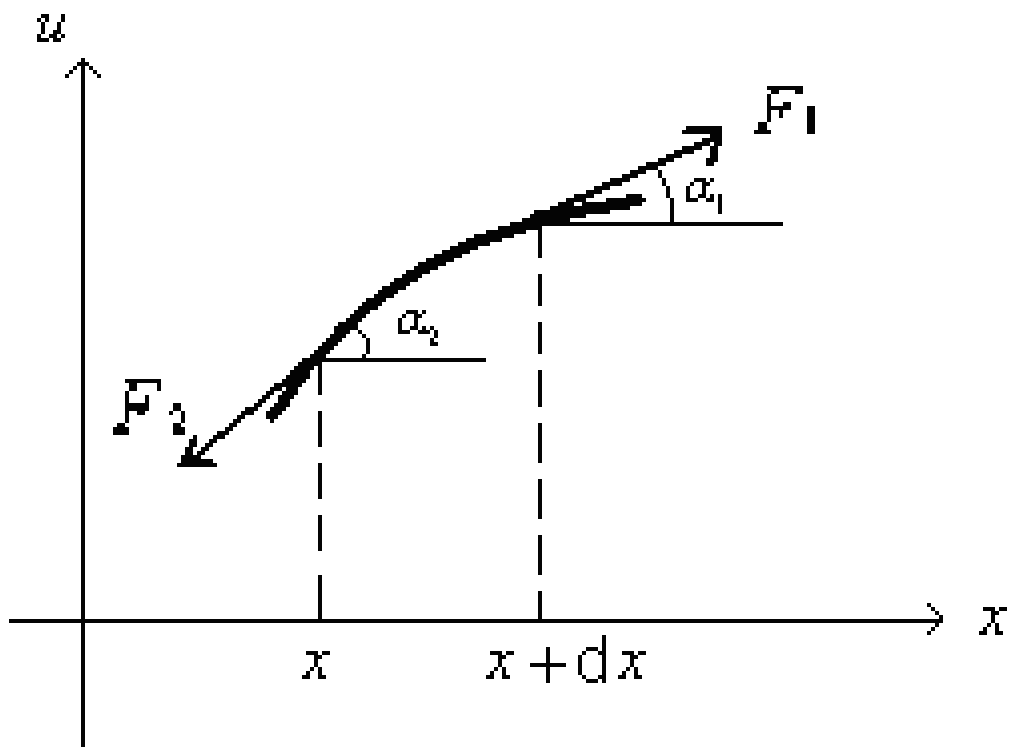

Fig 3.1. Analysis of forces on a string.

As shown in Fig. 3.1, we have

$$\alpha_1 \approx \alpha_2,\ F_1 \approx F_2 \equiv F\ ,$$

$$F_1 \sin\alpha_1 - F_2 \sin\alpha_2 = F_1 \left.\frac{\partial u}{\partial x}\right|_{x+\mathrm{d}x} - F_2 \left.\frac{\partial u}{\partial x}\right|_{x} = \rho \mathrm{d}x \frac{\partial^2 u}{\partial t^2}\ . \tag{3.1}$$

Expanding the left-hand side at $x$, we have

$$\frac{\partial^2 u}{\partial x^2} - \frac{\rho}{F}\frac{\partial^2 u}{\partial t^2} = 0 \tag{3.2}$$

Define $\frac{F}{\rho} = k^2$ and set $u(x,t) = X(x)T(t)$, we have two ordinary differential equations

$$\begin{cases} \dfrac{\mathrm{d}^2 T}{\mathrm{d}t^2} + \lambda T(t) = 0 \\ \dfrac{\mathrm{d}^2 X}{\mathrm{d}x^2} + \dfrac{\lambda}{k^2} X(x) = 0 \end{cases}, \tag{3.3}$$

where $\lambda$ is a constant to be determined.
Then

$$T_n(t) = A_n \sin(\sqrt{\lambda} t + B_n)\ , \tag{3.4}$$

$$X_n(x) = C_n \sin(\sqrt{\frac{\lambda}{k^2}}x + D_n), \tag{3.5}$$

$$u(x,t) = \sum_{n=1}^{\infty} X_n(x) T_n(t). \tag{3.6}$$

To fix $A_n, B_n, C_n$ and $D_n$, we need boundary and initial conditions. Set the effective length of the string to be $l$. The two ends of the string are fixed, so boundary conditions are:

$$X_n(0) = 0, X_n(l) = 0 \tag{3.7}$$

Thus we have

$$X_n(x) = C_n \sin(\frac{n\pi}{l} x), \tag{3.8}$$

$$T_n(t) = A_n \sin(\frac{kn\pi}{l} t + B_n) \tag{3.9}$$

Initial conditions are

$$u(x,0) = f(x), \quad \dot{u}(x,0) = v(x) = 0, \tag{3.10}$$

where

$$f(x) = \begin{cases} k_1 x, & (0 \le x \le c) \\ k_2(x-l), & (c \le x \le l) \end{cases} \tag{3.11}$$

where $x = c$ is the contact point of hammer and string, as shown in Fig. 3.2.

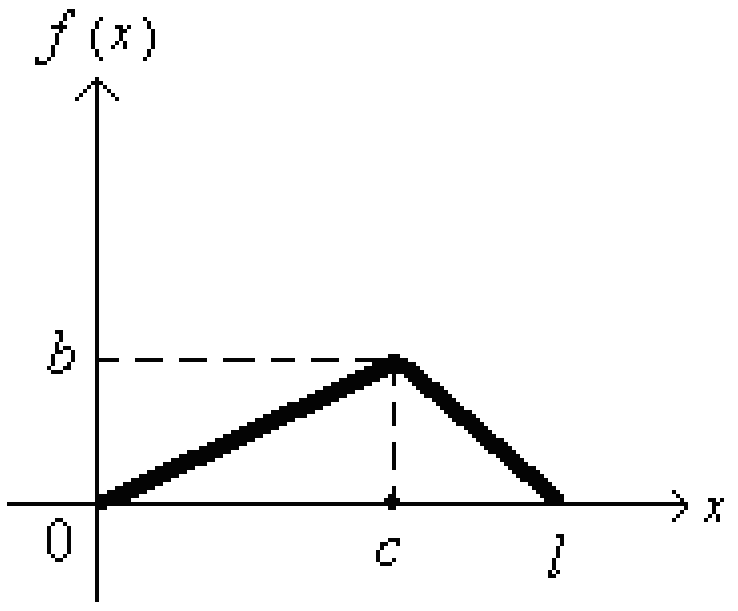

Fig.3.2. Initial conditions of string. $b$ is the amplitude of string vibration.

From $\dot{u}(x,0) = 0$, we know

$$\cos B_n = 0 \tag{3.12}$$

For simplicity, we set $B_n = \frac{\pi}{2}$, and then

$$
\begin{aligned}
u_n(x,t) &= A_n C_n \sin(\frac{n\pi}{l}x)\cos(\frac{kn\pi}{l}t) \\
&\equiv E_n \sin(\frac{n\pi}{l}x)\cos(\frac{kn\pi}{l}t)
\end{aligned}
\tag{3.13}
$$

From Fourier analysis, we have

$$
u(x,0) = f(x) = \sum_{n=1}^{\infty} E_n \sin(\frac{n\pi}{l}x)\,, \tag{3.14}
$$

$$
\begin{aligned}
E_n &= \frac{2}{l}\int_0^l f(x)\sin\frac{n\pi x}{l}\mathrm{d}x \\
&= \frac{2}{l}\left[\int_0^c k_1 x\sin\frac{n\pi x}{l}\mathrm{d}x + \int_c^l k_2(x-l)\sin\frac{n\pi x}{l}\mathrm{d}x\right] \\
&= \frac{2l^2}{n^2\pi^2 c}\cdot(-k_2)\cdot\sin\frac{n\pi x}{l} \qquad (k_2 < 0)
\end{aligned}
\tag{3.15}
$$

So $E_n$ is linear with $(-k_2) = |k_2|$, which means the ratio of harmonics remains the same when the string is hit more mightily; what is changed is only the loudness. But the experiments in Part 2 show clearly that the more strongly hit, the more higher frequencies the string will contain.

To reconcile this contradiction, I tried to find some other factors affecting the tones.

Firstly I thought of gravity, which might affect the tension in the vertically fixed string. Different sections experience different tensions, resulting in a new wave equation. Nevertheless, a grand piano’s strings are horizontally tightly fixed, and it still makes an enormous amount of tones. So the effect of gravity can be ignored.

Secondly I thought of the soundboard. The string’s vibrations cannot couple with the air around it very well; it is the bridge and soundboard that solve the problem. The former is used to transport energy of the strings’ vibrations to the latter, which radiates the vibrations into the air and makes the piano singing. Without the bridge and soundboard, the sound will be very weak.[3]

The importance of the soundboard led me to think that the effects of different pressing forces would ultimately be revealed by the motions of soundboard. However, a common phenomenon reminded me that the tone problem should not have deep relation with the soundboard, but with the hammer and the string: piano technicians’ tiny manipulations on a hammer or a string will cause an obvious change in the tone. So I ought to return to the hammer-string system.

## 4. Corrections of previous analysis, discussion and final conclusions.

Let's check the interaction between hammer and string again. The hammer hits the string with a small part of its arch contacting with the string. Then the corresponding part on the string has a velocity and leads the whole string to vibrate. Here we should pay attention to the initial conditions. The string is hit, not plucked like a guitar[4], i.e., the initial conditions in Eq. (3.10) are wrong. Instead, we have

$$u(x,0)=0 \quad \text{and} \quad \dot{u}(x,0)=v(x)\neq 0 \tag{4.1}$$

For simplicity, we take $v(x)=\begin{cases}0, & 0\le x\le c-d\\ b_0 & c-d\le x\le c+d\\ 0 & c+d\le x\le l\end{cases}$, with $[c-d,c+d]$ the contact range of hammer and string, as shown in Fig. 4.1.

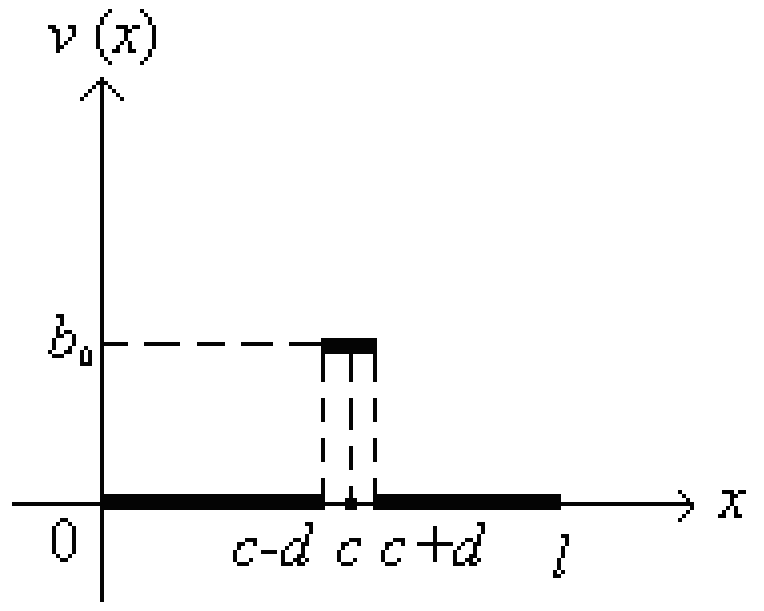

Fig. 4.1. Initial velocity of the string.

Using the new initial condition (4.1), we have

$$u_n(x,0)=E_n\sin\left(\frac{n\pi x}{l}\right)\sin\left(\frac{kn\pi t}{l}\right), \tag{4.2}$$

$$\dot{u}(x,0)=v(x)=\sum_{n=1}^{\infty}E_n\sin\left(\frac{n\pi x}{l}\right) \tag{4.3}$$

where $E_n\equiv A_nC_n$. Similar to previous analysis, we obtain $E_n$ from Fourier transform:

$$\begin{aligned}E_n&=\frac{2}{l}\int_0^l v(x)\sin\left(\frac{n\pi x}{l}\right)\mathrm{d}x\\ &=\frac{2}{l}\int_{c-d}^{c+d} b_0\sin\left(\frac{n\pi x}{l}\right)\mathrm{d}x\\ &=\frac{2b_0}{n\pi}\left\{\cos\left[\frac{n\pi}{l}(c-d)\right]-\cos\left[\frac{n\pi}{l}(c+d)\right]\right\}\end{aligned} \tag{4.4}$$

But disappointedly, we see $E_n$ is still linear with $b_0$, that is, when the key is hit more strongly, the ratios of the intensities of the frequencies on the string remain unchanged. So the magnitude of the string does not affect its tone, which contradicts with our experiment in Part 2.

I analyzed the coefficient $E_n$ in Eq. (4.3). When the key is under a bigger force, which parameters will change? Obviously $b_0$ will increase; more importantly, I finally noticed that $d$ would increase, too. Since the hammer is made of wool felt, it has some elasticity. So when it hits the string with a bigger velocity, the contact range should expand, that is, $d$ increases.

First, I tried $E_1$ and $E_2$ to calculate whether the increment of $d$ will cause a brighter sound. Initial velocities are shown in Table 4.1.

| Press force | Small | Big |
|---|---|---|
| $E_n$ | $E_1, E_2$ | $E_1', E_2'$ |
| Contact range between hammer and string | $2d$ | $2d'$ |
| $\dot{u}(x,0) = v(x)$ | 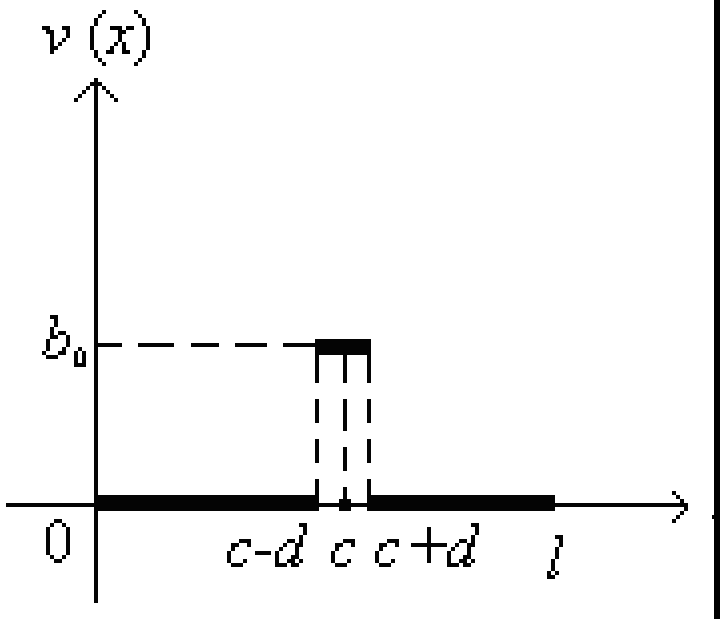  | 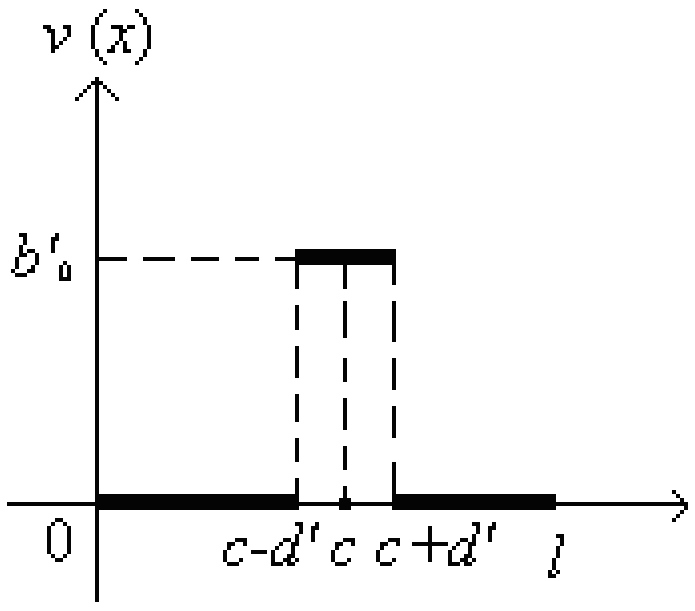  |

Table 4.1 Comparison of different initial conditions due to different press forces

We expect $\frac{E_1}{E_2} > \frac{E_1^{'}}{E_2^{'}}$, i.e.

$$\frac{\cos\frac{\pi(c-d)}{l} - \cos\frac{\pi(c+d)}{l}}{\cos\frac{2\pi(c-d)}{l} - \cos\frac{2\pi(c+d)}{l}} > \frac{\cos\frac{\pi(c-d')}{l} - \cos\frac{\pi(c+d')}{l}}{\cos\frac{2\pi(c-d')}{l} - \cos\frac{2\pi(c+d')}{l}} \tag{4.5}$$

Using the identity

$$\cos(x_1 - x_2) - \cos(x_1 + x_2) = 2\sin x_1 \sin x_2,$$

we only need to see whether

$$\sin\frac{\pi d}{l}\sin\frac{2\pi d'}{l} > \sin\frac{\pi d'}{l}\sin\frac{2\pi d}{l}, \tag{4.6}$$

i.e.,

$$\cos\frac{\pi d'}{l} > \cos\frac{\pi d}{l} \tag{4.7}$$

For an upright piano, the strings of its middle range are at the length around 0.4m~0.5m, and the contact length between hammer and string is less than 1cm. Thus

$$0 < \frac{\pi d}{l} < \frac{\pi d'}{l} < \frac{\pi}{2}. \tag{4.8}$$

Then the inequality of (4.7) is wrong. On the contrary, we have

$$\cos\frac{\pi d'}{l} < \cos\frac{\pi d}{l}, \quad \frac{E_1}{E_2} < \frac{E_1^{'}}{E_2^{'}}. \tag{4.9}$$

That is, the intensity of first harmonic frequency $f_2$ increases slower than the fundamental when the press force increases, leading to a softer tone, which is contradictory with the experimental result.

For common cases of $d' > d$ and $n > m$, we can also derive that $\frac{E_m}{E_n} < \frac{E_m^{'}}{E_n^{'}}$ (see Appendix).

This leads us to a dilemma: either the experiment in Part 2 has something wrong or the theoretical analysis fails somewhere.

After examining the whole experiment, I turned to check the analyzing process, especially the interaction between the hammer and string. Besides the amplitude of the string and the contact range between hammer and string, there is another important variable, the contact time $t_c$. What will happen to $t_c$, if the force exerted on the key increases?

There are several possible relevant factors (see Part 5: Discussions of unsolved problems). Here we study one of them, the elasticity of the hammer.

A felt-covered hammer is considered as a spring with stiffness coefficient $k_0$. Set the mass of the hammer to be $m$. Since the string's amplitudes are very small, we can treat the string as a rigid body when it interacts with the hammer.

The equation of the hammer's motion is

$$m\frac{\mathrm{d}^2 X}{\mathrm{d}t^2} = -k_0 X(t), \tag{4.10}$$

with its solution

$$X(t) = A\left(\sin\sqrt{\frac{k_0}{m}}t + B\right). \tag{4.11}$$

The contact time is

$$t_c = \frac{T}{2} = \frac{2\pi}{2\sqrt{\frac{k_0}{m}}} = \pi\sqrt{\frac{m}{k_0}}\ . \tag{4.12}$$

In fact, $k_0$ is not a constant. It varies with the displacement $X$ because of the manufacturing process of the hammer. In detail, $k_0$ increases nonlinearly from the outside to the inside of the hammer.[5] So Eq. (4.10) becomes

$$m\frac{\mathrm{d}^2 X}{\mathrm{d}t^2} = -k_0(X)X(t)\,. \tag{4.13}$$

We don’t have to solve this equation, since what we want is the comparison of the contact time $t_{c1}$ and $t_{c2}$ under different press forces $F_1$ and $F_2$ respectively ($F_1 < F_2$). Instead, we can take the average value of $k_{0(1)}$ and $k_{0(2)}$ (corresponding to $F_1$ and $F_2$). Since

$$F_1 < F_2\ ,\quad k_{0(1)} < k_{0(2)}\,,$$

qualitatively we know from Eq. (4.12) that

$$t_{c1} > t_{c2}\,, \tag{4.14}$$

namely, when the knocking force on the key increases, the contact time between hammer and string decreases.

Now we need to compare the periods $T_n$ of vibrations on the string with the contact time $t_c$ when hammer strikes the string. If $t_c << T_n (1 \le n \le 6)$, the contact of hammer won’t affect the string’s vibrations much. If $t_c$ and $T_n$ are at the same order of magnitude or $t_c >> T_n$, then the hammer will depress the vibrations on the string for a while, which means the tone may be influenced.

From Eq. (4.2) we know

$$T_n = \frac{2\pi l}{kn\pi} = \frac{2l}{kn}\,, \tag{4.15}$$

where $k = \sqrt{\frac{F}{\rho}}$ is the wave speed.

Here are some data to help us estimate $k$[5]:

In the middle register of an upright piano:

A string's linear density: $\rho \approx 6\text{g/m}$.

The tension in the string: $F \approx 1800\text{N}$.

Thus

$$k \approx \sqrt{\frac{1800\text{N}}{6\times10^{-3}\text{kg/m}}} = 547.7\text{m/s} \tag{4.16}$$

Then for $n = 1$ and $l \approx 0.4\text{m}$, we have:

$$T_1 = \frac{2l}{kn} \approx 1.5\times10^{-3}\text{s}. \tag{4.17}$$

Having had $T_1$, we now would like to know the contact time $t_c$. The $t_c$ is so short that it is difficult to be measured through common experiments. Lacking professional equipment to measure $t_c$, we'll estimate it through theoretical analysis. Here are some data:

The mass of a hammer in the middle register[6]: $m \approx 8\text{g}$.

The distance between the hammer and string when the key is released: $d_h \approx 5\text{cm}$.

Since the string's amplitudes are very small, it can be treated as a rigid body when hit by the hammer.

Set the average value of the force between hammer and string to be $\bar{F}_h$, and the speed at which the hammer heads forwards and goes back to be $v_h$. Therefore,

$$\bar{F}_h \cdot t_c = 2mv_h, \tag{4.18}$$

then

$$\bar{F}_h = \frac{2mv_h}{t_c}. \tag{4.19}$$

We now estimate $v_h$. A complete striking process is divided into three sections, as shown in Table 4.2:

| Section | Time needed |
|---|---|
| I The hammer starts to move until it contacts with the string | $t_1$ |
| II The hammer interacts with the string | $t_c$ |
| III The hammer leaves the string and goes back to its initial position | $t_1 = t_2$ |

Table 4.2 Three sections of a complete striking process of hammer to string

A single key of a well-made upright piano can be knocked down 8 times per second.[5] Thus

$$t_1 \approx \frac{1-8t_c}{8} \cdot \frac{1}{2}, \tag{4.20}$$

$$v_h \approx \frac{d_h}{t_1} = \frac{16d_h}{1-8t_c}. \tag{4.21}$$

Then

$$\bar{F}_h = \frac{2m \cdot 16d_h}{(1-8t_c)t_c}. \tag{4.22}$$

Let's see what will happen when $t_c << T_n$. Taking $t_c \approx 10^{-4}\,\text{s}$ as a trial, we get $\bar{F}_h \approx 120\,\text{N}$.

However, practically the hammer will experience a force about 10N~20N when hitting the string.[6][7] Here the force with its average magnitude 120N is supposed to damage the hammer.[5] So the contact time between hammer and string should be more than $10^{-4}\,\text{s}$. Comparing it with the periods of standing waves on the string:

$$T_1 \approx 1.5\times10^{-3}\,\text{s}, \quad T_2 \approx 7.5\times10^{-4}\,\text{s}, \quad T_3 \approx 3.75\times10^{-4}\,\text{s} \quad \text{and so on,}$$

we know that $t_c$ is at least at the same order of magnitude with $T_1$. There are experimental proofs of this estimation of contact time in Reference [7] and in the "Introduction" of Reference [6]: "…we are not surprised to see that measured contact times in the middle range are quite similar to half of the period with which the string would vibrate alone, and longer than the vibration periods of most the upper harmonics."

If $t_c$ increases due to decrement of press force on the key, the wave with higher frequencies on the string are expected to lose more energy to the hammer, because they undergo more periods during the contact. Then the fundamental or lower frequencies, dominates in the vibration, causing the tone to be softer.

We now tabulate the discussion above:

| The magnitude of the press force | Small | Big |
|---|---|---|
| Contact time between hammer and string ($t_c$) | Longer | Shorter |
| The ratio: $\frac{E_{high}}{E_{whole}}$<br>$E_{high}$: energy of higher frequencies absorbed by the hammer<br>$E_{whole}$: energy of all frequencies absorbed by the hammer | Big | Small |
| The tone | Brighter and sharper | Softer |

Table 4.3 Brief summary of previous discussion

So far we've corrected the physical model of a piano string's vibration from plucking to knocking, and used the latter to draw several conclusions:

① The amplitude of string doesn't affect the tone.
② The increment of contact range between hammer and string (due to the increment of force on the key) causes the tone to become softer (more details in Appendix).
③ The decrement of contact time between hammer and string (due to the increment of force on the key) causes the tone to become brighter.

## 5. Discussion of unsolved problems

(I) In Part 4 we obtained the conclusion that when the magnitude of the force increases, the contact range and contact time between hammer and string contribute inversely to the tone. According to the experiments in Part 2, we know the factor of contact time should dominate.

We have tried a simple case that $\dot{u}(x,0)=v(x)$ is a constant $b_0$ within the contact range [ $c-d$ , $c+d$ ]. In fact, $v(x)$ should be more complex and it is difficult to determine it. At present we cannot decide accurately how the hammer hits the string unless we measure it through appropriate experiments.

In fact, the contact time is affected not only by the changing elasticity of the hammer, but also some other factors, such as the mechanical parts connecting the key and the hammer. The analysis and calculation around the mechanical structure is so complicated that it can hardly be carried out by hand. Computer simulation should be a good alternative.

(II) From the very beginning of our discussion, i.e., the derivation of wave equation on the string, we admit "tacitly" that the string is completely soft. But in fact, the piano string has some rigidity which causes dispersion.[8] We should know how "hard" a piano string is and add rigidity to the wave equation.

(III) We have postulated that the two ends of the effective part of a string are fixed almost perfectly. It is so at the tuning pins, but it is not perfectly fixed at the bridge. More accurate analysis should include the vibration (even though very small) of the bridge and soundboard.

(IV) The problem about the tone of a piano is an interesting question, in the sense of both physics and psychology. On one hand, what we have done will never replace the technician's techniques and the listener's ears. On the other hand, we still can do more in the theoretical research of simulation and analysis of real pianos, to make piano manufacturing and playing more accurate and pleasant.

## 6. Acknowledgements

During the process of researching and accomplishing this course thesis, I gained a lot of directions and advices from Prof. Xi Chen, Prof. Dexin Lu and Prof. Yu An. I would like to express my sincere appreciation to them here.

## 7. Appendix

Here are some mathematical details in Part 4.

Common cases:

| Press force | Small | Big |
|---|---|---|
| Coefficients ($n>m$) | $E_m, E_n$ | $E_m^{'}, E_n^{'}$ |
| Contact range between hammer and string ( $d'>d$ ) | $2d$ | $2d'$ |
| $\dot{u}(x,0)=v(x)$ | 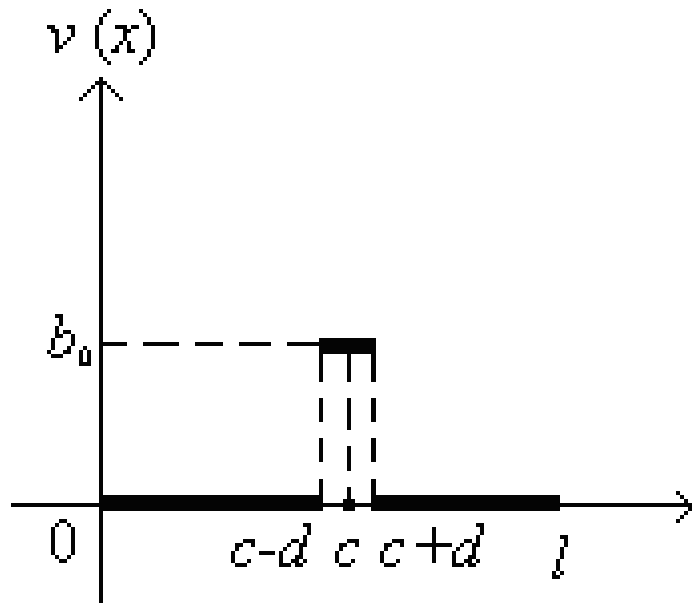  | 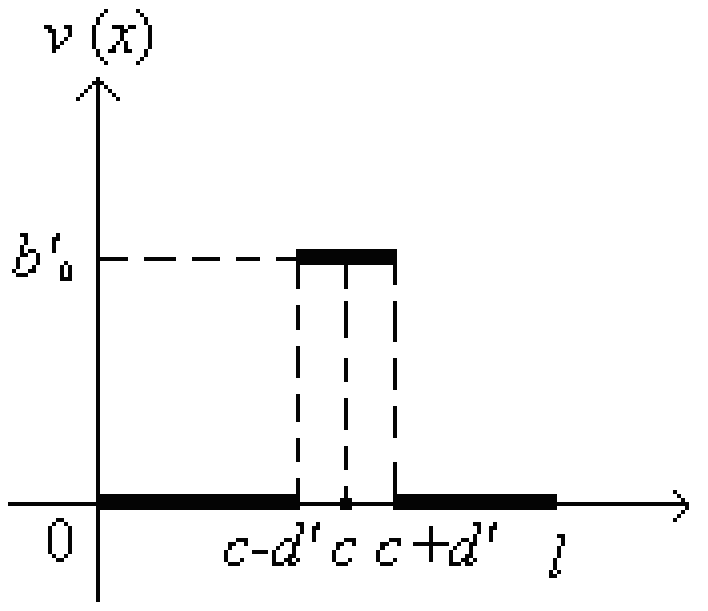  |

Table 7.1 Comparison of different initial conditions due to different press forces

We'll prove that

$$\frac{E_n}{E_m} > \frac{E_n^{'}}{E_m^{'}} . \tag{Ap.1}$$

Even though this goes against the experimental result in Part 2, it can offer an explanation to a problem in piano adjustment, which is used to be solved empirically.

We'll prove that

$$\frac{\dfrac{2b_0}{n\pi}\left[\cos\dfrac{n\pi(c-d)}{l}-\cos\dfrac{n\pi(c+d)}{l}\right]}{\dfrac{2b_0}{m\pi}\left[\cos\dfrac{m\pi(c-d)}{l}-\cos\dfrac{m\pi(c+d)}{l}\right]} > \frac{\dfrac{2b_0'}{n\pi}\left[\cos\dfrac{n\pi(c-d')}{l}-\cos\dfrac{n\pi(c+d')}{l}\right]}{\dfrac{2b_0'}{m\pi}\left[\cos\dfrac{m\pi(c-d')}{l}-\cos\dfrac{m\pi(c+d')}{l}\right]},$$

i.e.,

$$\sin\frac{n\pi d'}{l}\sin\frac{m\pi d}{l} < \sin\frac{n\pi d}{l}\sin\frac{m\pi d'}{l} \tag{Ap.2}$$

Now we would like to use Taylor expansion to check whether Eq. (Ap.2) is correct. In an upright piano, the effective length of a string in the middle register is about 0.4m~0.5m, and contact range $2d$ is about 0.5cm~1cm. If we set $n \leqslant 6$, we'll get

$$\left(\frac{n\pi d}{l}\right)^3 << 1 \tag{Ap.3}$$

So (Ap.2) becomes

$$\left[\frac{n\pi d'}{l}-\left(\frac{n\pi d'}{l}\right)^3\cdot\frac{1}{3!}\right]\left[\frac{m\pi d}{l}-\left(\frac{m\pi d}{l}\right)^3\cdot\frac{1}{3!}\right] < \left[\frac{n\pi d}{l}-\left(\frac{n\pi d}{l}\right)^3\cdot\frac{1}{3!}\right]\left[\frac{m\pi d'}{l}-\left(\frac{m\pi d'}{l}\right)^3\cdot\frac{1}{3!}\right]$$

i.e.,

$$\frac{n^2\pi^2}{l^2}\left(d'^2-d^2\right) > \frac{m^2\pi^2}{l^2}\left(d'^2-d^2\right) \tag{Ap.4}$$

Since $n > m$ , $d' > d$ , (Ap.4) is correct. So is (Ap.1).

Now we have the conclusion about the effect of the contact range between hammer and string: the bigger it is, the softer the tone will be.

This can be used to explain a phenomenon in piano adjustment. When the piano's sound seems too bright and sharp, piano technicians usually use emery paper to grind the hammer. When the head of the hammer becomes appropriately flat, the tone will be softened. Sometimes over-grinding will cause the tone to be dark and dim. One of the keys of my own piano unfortunately underwent this kind of dim voice because of over-grinding on the hammer.